%% file: 0.main.tex
\documentclass[a4paper,11pt]{article}
\usepackage{authblk}
\usepackage{cite}
\usepackage{amsmath,amssymb,amsfonts}
\usepackage[ruled,vlined]{algorithm2e}
\usepackage{algpseudocode}
\usepackage{graphicx}
\usepackage{textcomp}
\usepackage{xcolor}
\usepackage{soul}
\def\BibTeX{{\rm B\kern-.05em{\sc i\kern-.025em b}\kern-.08em
    T\kern-.1667em\lower.7ex\hbox{E}\kern-.125emX}}

\providecommand{\keywords}[1]
{
  \small	
  \textbf{\textit{Keywords---}} #1
}

\begin{document}

\title{CASH: A Credit Aware Scheduling for Public Cloud Platforms}

\author[1]{Aakash Sharma\thanks{abs5688@psu.edu}}
\author[2]{Saravanan Dhakshinamurthy\thanks{saravanand@fb.com}}
\author[1]{George Kesidis\thanks{gik2@psu.edu}}
\author[1]{Chita R. Das\thanks{cxd12@psu.edu}}
\affil[1]{Department of Computer Science and Engineering, The Pennsylvania State University, University Park}
\affil[2]{Facebook Inc.}

\renewcommand\Authands{ and }
\date{}
\maketitle

\begin{abstract}
The public cloud offers a myriad of services which allows its tenants to process large scale big data in a flexible, easy and cost effective manner. Tenants generally use large scale data processing frameworks such as MapReduce, Tez, Spark etc. to process their data. Tenants can configure their frameworks to run individual tasks by the framework itself or have a middleware cluster manager like YARN or Mesos to arbitrate resource scheduling in their public-cloud cluster. Cluster managers need to be cognizant about the workload requirement along with the state of the individual resource such as CPU and disk in the cluster. Cloud providers use a token bucket mechanism for their individual hardware resources as an indicator of the quality-of-service that individual hardware resource can provide. In this paper, through our changes in YARN, Hadoop and Tez, we show how middleware cluster managers can be made cognizant about the expected quality-of-service of individual hardware resources in the cluster. Our optimized cluster manager with a coarse grained knowledge of task requirement and fine grained knowledge of expected quality-of-service of hardware resources in the cluster performs highly optimal task placements. Our experiments with our optimizations show CPU credit based instances like the Amazon T3 instances as a viable cost effective option for running big data workloads. We also show that streaming SQL queries on a Hive warehouse can be accelerated by up to 31\% leading to public cloud billing cost savings of up to 22\%. 

\end{abstract}


\input{1.introduction}

\input{2.background}

\input{3.motivation}

\input{4.methodology}

\input{5.implementation}
\input{6.evaluation}

\input{7.relatedwork}
\input{8.conclusion}

 \bibliographystyle{IEEEtran}

\bibliography{references,cloud,scheduling,DDoS}

\end{document}

%% file: 1.introduction.tex
\section{Introduction} \label{sections: introduction}

Big data workloads are processed using a compute cluster wherein the data set to be processed is either distributed over a large number of storage volumes or is resident in a storage service offering capacity in the order of at least a few TBs.
These large data sets typically require a large amount of compute resources to be processed which is easily made available in a cloud setup. 
This work focuses on running big data workloads on a public cloud setup in a cost effective manner.\par

Public cloud computing has become a ubiquitous part of every big data processing entity owing to its flexibility and cost savings over private data centers. 
Public cloud offerings 
include enterprise services spanning compute, storage and networking. 
Public cloud providers allow users to lease hardware and software services of desired capacity over fixed or variable time periods. 
So, tenants can lease only the amount of resources required to meet their workload demands. 
In a traditional private data center, tenants pay the total cost of ownership (CapEx and OpEx) of the data center for its entire lifetime. 
Any unused capacity at any given point of time translates to losses for a tenant. On the flip side, workloads requiring more resources than the current capacity of the private data center may lead to violations of Service-Level Objectives (SLOs). 
While a tenant may not need the peak capacity needed to meet a given SLO, they nonetheless may need to maintain hardware resources at peak capacities to honor all SLOs.
This can be avoided in a public cloud as tenants can dynamically choose the scale of their resources based on their workload demands at a given point in time leading to significant cost savings.

\begin{figure}
    \centering
    \includegraphics[width=\linewidth]{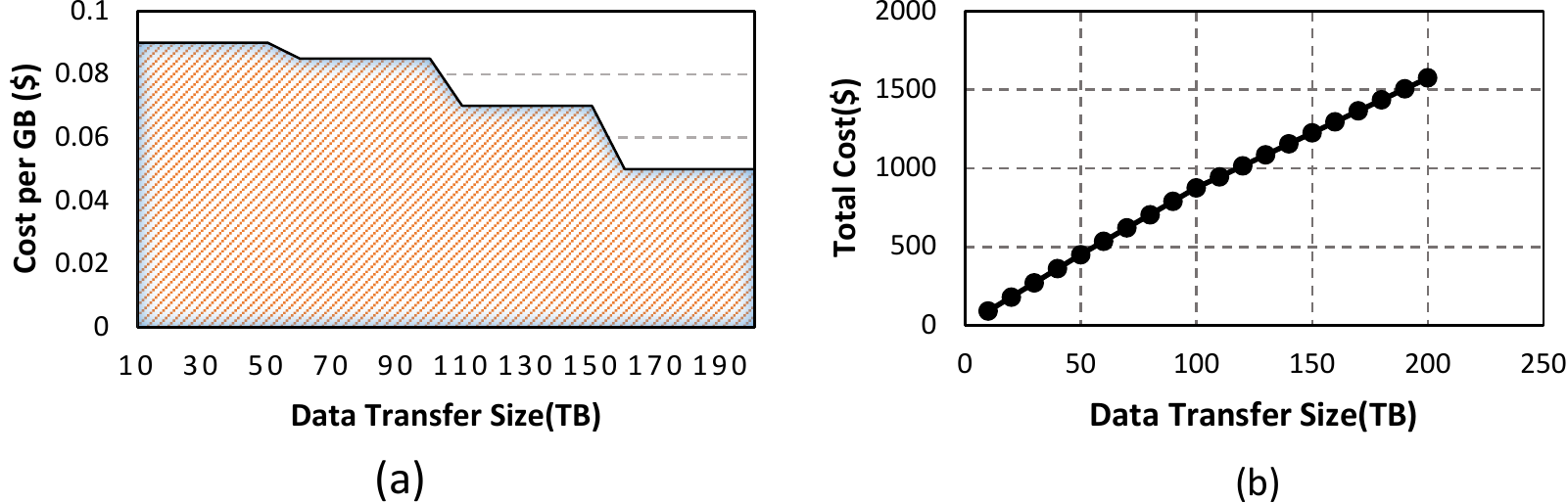}
    \caption{S3 Pricing: (a) Cost per GB for different sizes of data transfer; (b) Total cost of transferring data}
    \label{fig:s3_pricing}
\end{figure}

The other major contributor to the rise in popularity of the public cloud is low cost storage solutions provided by the public cloud. Services such as Amazon S3 \cite{s3} and  Azure blob storage \cite{azure-blob} are highly reliable and accessible storage solutions for an enterprise.
Tenants can store any arbitrary volume of data on these storage services and be billed for the exact volume of data resident on these services at any given point in time, without the need for \textit{pre-provisioning} of any storage volume. 
This leads to flexibility in storage planning and cost savings owing to being billed on the granularity of data volume rather than the underlying storage hardware. 
However, most cloud providers bill their tenants for outgoing data transfer from their storage services, with cost being dependent on the destination of the transfer. These data transfer costs are typically significantly more when the data transfer destination is outside of the cloud provider's premises than when it is within. As an example, the cost of transferring data from Amazon S3 to the internet is given in Figure \ref{fig:s3_pricing}. Compared to this, S3 charges between \$0.00--\$0.02 per GB when the destination is within AWS premises.
Since the cost of transferring a few TBs of big data out of AWS S3 can be quite high, tenants generally prefer processing their big data workloads within their public cloud provider premises.\par

MapReduce \cite{mapreduce}, Spark \cite{spark}, Flink \cite{apache-flink} and Tez \cite{Tez} are some of the popular frameworks for processing large-scale big data. 
These frameworks work by slicing a particular job into various small microtasks \cite{microservice} which may be grouped together for execution based on common characteristics. 
The framework then creates an execution plan for performing the given job which is basically an ordering of the execution of its various tasks. 
The framework can execute tasks by itself using a job scheduler built into it or submit the task for execution to an external job scheduler. 

A middleware cluster manager, e.g.,
Kubernetes \cite{kube-scheduler}, Mesos \cite{Mesos}
or YARN \cite{yarn,spark-vs-tez}, serves as a  resource arbitrator, i.e., it offers resources to a framework for its task execution. 
The scheduling algorithms employed in cluster managers typically consider every hardware resource in the cluster in an (often crude) attempt to provide a fixed service rate.
However, the public cloud offers a variable service rate because its hardware resources may be shared. Typically, memory and CPU resources are considered for scheduling decisions with the assumption that a ``CPU core" is constant. With variable service-rate CPU cores in the public cloud, a ``core" in the cluster will deliver variable service rate of up to 100\%. This variability along with other variable service rate resources like disk I/O and network I/O leads to sub-optimal scheduling decisions.
While disk and network I/O resources are typically not taken into consideration for making scheduling decisions, tasks may be  allocated on a critical (for it) hardware resource whose service rate is throttled.

Public cloud providers expose variability in service rate through Service-level-agreements(SLAs). 
Providers such as AWS\cite{aws} and Google cloud \cite{gcloud} quantify their variable rate hardware offerings and tenants can use this quantification to determine the expected service rate of resources in these instances. 
In case of AWS, token-buckets is used to regulate the availability of a resource (and hence the service rate it is expected to provide). Tenants can actively look-up the state of the tokens associated with a particular IT resource to determine the expected service rate. For example, CPU resources belonging to AWS T3\cite{t3} burstable instances\footnote{a.k.a. bursting, credit based or token based instances} and disk volumes belonging to AWS Elastic Block Storage (EBS) \cite{ebs}\footnote{EBS tokens are used even for non-bursable AWS instances.} have token buckets associated with them whose states can be dynamically queried using publicly available APIs.

We propose a novel credit aware scheduler -- CASH for a  cluster manager that considers both token-bucket state information associated with its VMs and the (only roughly disclosed) critical resource needs of the tasks it is scheduling. 
We prototype CASH on YARN, i.e., it queries the cloud provider for token-bucket state of its nodes and receives additional task information from its Tez and Hadoop application frameworks. 
As examples, we run both batch workloads and streaming SQL queries which are CPU and disk I/O intensive, respectively.
\\
\\
\\
\\
In summary, CASH makes the following key contributions:
\begin{itemize}
\item The efficacy of using token bucket information associated with hardware resources in task placement decisions leading to cost savings and task acceleration.
\item Empirically proving T3 burstable instances to be a viable and economical solution to running batch based big data workload.
\item Accelerating streaming SQL queries running over in-memory DAG based frameworks and reducing their job completion time by an average of 31\% which translates to an overall workload completion time shortening of 22\%.
\end{itemize}

This paper is organized as follows.
In section \ref{sections: background}, we provide some background discussion. In section \ref{sections: motivation}, we motivate our problem in detail, including through experimental case studies.
Our scheduling scheme is described in section \ref{sections: scheme} and how it was prototyped on Tez/Hadoop-over-YARN is described in section \ref{sections: implementation}.  The results of our experimental performance evaluations are given in section \ref{sections: evaluation}. In section \ref{sections: relatedwork} we discuss related work. Finally, we summarize in section \ref{sections: summary} and discuss future work.

%% file: 2.background.tex
\section{Background} \label{sections: background}

In this section, we discuss some service offerings by the world's largest public cloud provider -- AWS, which are relevant to this paper. 
Other large public cloud providers (particularly 
Microsoft Azure and Google Cloud) offer similar services.

\subsection{AWS T3 burstable instances}
AWS T3s are CPU credit based instances that have a guaranteed baseline CPU service rate which is a percentage of an actual VM CPU core of a comparable general purpose type. 
The VM acquires CPU credits when operating below the ``baseline" service rate up to the maximum size of the associated token bucket. 
Each credit can be used to ``burst" to 100\% CPU for one minute or 50\% CPU for 2 minutes. 
The amount of credits an instance acquires is at millisecond granularity and can be tracked using APIs provided by Amazon. 
The rate of credit earned is dependent on the instance size. 
Table~\ref{table:t3credits} provides a few T3 instance sizes, their configurations and their CPU credit properties from the AWS website\cite{t3}.
T3 also supports an unlimited credit option which prevents tenants from being throttled to baseline performance if they run out of credits. The credit balance is calculated as an average over 24 hours, or the instance lifetime, whichever is shorter, and tenants are billed for the excess credits they use. 
\input{Main/Tables/t3_credits.tex}

\subsection{AWS Elastic Block Storage (EBS)}
\begin{figure}
    \centering
    \includegraphics[width=.5\linewidth]{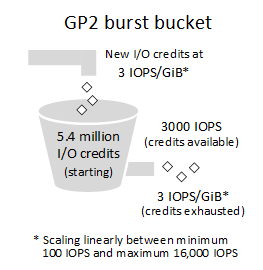}
    \caption{EBS token bucket}
    \label{fig:ebsTokenBucket}
\end{figure}

AWS EBS is a block storage service which can be attached to any EC2 instance. 
EBS storage volumes are persistent volumes and their lifetime is independent of the AWS instance they are attached to. 
The size of EBS volumes is specified by its user upon creation and can range from a single GB to 16 TB. EBS volumes fall into two main categories -- SSD backed and HDD backed. 
The performance of the volume increases with the size of the volume provisioned, with SSD backed volumes delivering significantly higher IOPS than HDD backed volume for a little more than double the price per GB. 
However, performance and service rate of EBS volumes is variable like other public cloud offerings with EBS using a similar token bucket mechanism as T3 instances. 
EBS SSD backed volumes have a baseline Input/Output per second (IOPS) credit rate of three times their size in GB. 
When the volume uses less IOPS than its baseline credit arrival rate, credits are conserved in the corresponding token bucket. 
Volumes can use their accumulated I/O credits to burst to a maximum performance of 3000 IOPS. 
A screen capture from AWS \cite{ebs_token_bucket} describing the EBS token bucket is given in Figure \ref{fig:ebsTokenBucket}.

\subsection{Amazon Elastic MapReduce (EMR)}
Amazon EMR\cite{emr} is essentially a big data processing SaaS offering from AWS which allows tenants to run application frameworks like Hadoop, Spark, Tez etc. along with databases such as Hive\cite{hive} to process big data.
EMR can be provisioned within minutes and tenants can have a running data processing cluster to run their workloads. 
EMR comes with the YARN capacity scheduler as the cluster manager. 
Users choose the type of hardware they want to use to create their cluster. 
Users specify the instance type and the type of storage that needs to be attached. 
A key restriction of EMR however, is that it does not allow users to create their cluster with low cost credit based T3 burstable instances \cite{emrInstances}.

\subsection{Amazon S3}
Amazon S3\cite{s3} is an object based storage service offering by Amazon which offers cheap and reliable storage service. 
Tenants of S3 can expect to receive 99.999999999\% durability for their data \cite{s3} along with flexibility to store data of any size and access it from anywhere over the internet.
Tenants running large-scale big data batch workloads on the public cloud prefer using object stores such as AWS S3 as their data storage for reasons explained in section \ref{sections: introduction}. 
A Nasdaq case study \cite{nasdaq} further explains this preference.
Most importantly, object stores provides the cheapest storage option in the public cloud. 
On AWS, while SSD backed volume costs \$0.10 per GB and a HDD backed volume costs \$0.045 per GB, S3 costs only \$0.023 per GB for the first 50 TB with a \$0.0004-\$0.005 per API call charge to data objects.

%% file: Main/Tables/t3_credits.tex
\begin{table}[]
\begin{center}

\begin{tabular}{|l|l|l|l|l|}
\hline
\textbf{Type} & \textbf{vCPUs} & \textbf{\begin{tabular}[c]{@{}l@{}}Memory\\  (GiB)\end{tabular}} & \textbf{\begin{tabular}[c]{@{}l@{}}Baseline \\ Performance/vCPU\end{tabular}} & \textbf{\begin{tabular}[c]{@{}l@{}}CPU credits \\ earned / hr\end{tabular}} \\ \hline
t3.large      & 2              & 8                                                                & 30\%                                                                          & 36                                                                          \\ \hline
t3.xlarge     & 4              & 16                                                               & 40\%                                                                          & 96                                                                          \\ \hline
t3.2xlarge    & 8              & 32                                                               & 40\%                                                                          & 192                                                                         \\ \hline
\end{tabular}
\end{center}

\caption{AWS T3 CPU credits}
\label{table:t3credits}
\end{table}

%% file: 3.motivation.tex
\section{Motivation} \label{sections: motivation}

\subsection{CPU Utilization in Compute Clusters}
Large compute clusters commonly face low resource utilization and efficiency even after collocating online services and batch workloads. 
Analysis\cite{Alibaba_17_Imbalance,Alibaba_17_workloadLook,Alibaba_18_resourceEfficiency} done on recently released Alibaba production traces \cite{alibabaTrace} shows that average CPU utilization remains below 40\% for about 60\% of the machines and below 50\% for 75\% of the machines in their cluster. 
However, the CPU usage follows a bursty pattern with CPU utilization going above 60\% for only about an hour in a 24 hour window for majority of the machines. 
This should prevent Alibaba from using fewer CPUs to run their workloads as it would violate their SLAs due to transient increases in CPU demands. 
Analysis of the Taoboa Hadoop cluster \cite{Ren} also found CPU utilization to not exceed 40\%. 
Even the Google trace analysis\cite{Reiss} reveals that CPU utilization doesn't exceed 60\%, where the  trace comprised of a mix of long running services, MapReduce and HPC.
An additional factor which contributes to low CPU utilization in a public cloud is a preference to use cheap object based external storage over traditional locally available distributed storage system like HDFS as already discussed.

\begin{figure}
    \centering
    \includegraphics[width=1\linewidth]{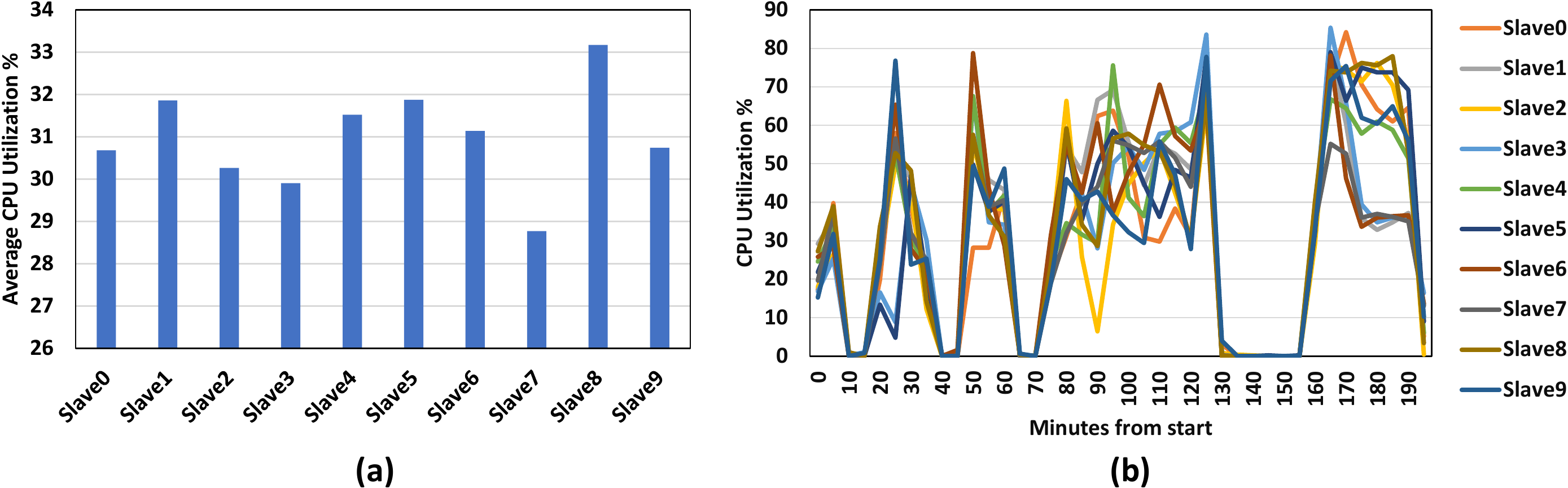}
    \caption{CPU utilization in EMR: (a) Average utilization per VM, (b) CPU utilization timeline}
    \label{fig:CpuEMRNode}
\end{figure}

\subsubsection{CPU utilization on EMR}
We tested MapReduce performance on EMR using the same workloads we use later for evaluation. We found a low CPU utilization with the average of about 30\% per node as given in Figure \ref{fig:CpuEMRNode}(a). This is due to high object read/write latency of S3 \cite{pelleS3Latency}. The timeline of CPU utilization on EMR is given in Figure \ref{fig:CpuEMRNode}(b).
One might be tempted to reduce the number of instances to improve CPU utilization, but this will further degrade I/O performance as there will be less parallelism during read/write to S3.

\subsubsection{The case for CPU credit based Instances}
Big data batch workloads that have low \emph{average} CPU utilization can benefit from the low cost offering of burstable instances. 
Table~\ref{table:t3PriceComparison} compares the pricing of AWS T3 with equivalent general purpose AWS M5 instances and AWS EMR.
\input{Main/Tables/t3PriceComparison.tex}
As we can see, (regular) M5 instances are more than 15\% more expensive than T3 instances and EMR on M5 is more than 44\% more expensive than on (burstable) T3. 
Tenants can obtain significant cost savings by using T3 (including for MapReduce workloads) over regular VMs with the same
peak IT resource allocations. 
For example, MapReduce workloads are often low CPU utilization workloads and they can particularly benefit from low cost, low CPU throughput instances like T3. This said, running Hadoop naively on T3 resulted in as much as 40\% degradation in elapsed task time with our workloads and hence is not the most optimal solution. This is captured in our evaluation.

\subsection{I/O Utilization in Compute Clusters}
An analysis of TPC-DS workloads 
in 
\cite{tpc_ds_analysis_socc18} shows very high I/O utilization (up to 100\%) for their big-data systems based on HDFS. 
Anecdotal evidence suggests that the disk is typically a significant bottleneck in I/O intensive workloads and although earlier work such as \cite{making_sense_performance_nsdi15} observe that CPU may also be a significant bottleneck, they also note that this is due to CPU time spent in compression/decompression of data to/from disk and serialization/de-serialization of java objects from/to byte buffer. 
Since the experiments done in \cite{making_sense_performance_nsdi15} were with  three generations older architecture and newer results show high I/O utilization in current production clusters, we can assume that disk I/O may again be a bottleneck for running I/O intensive SQL queries.\par
\begin{figure}
    \centering
     \includegraphics[width=1\linewidth]{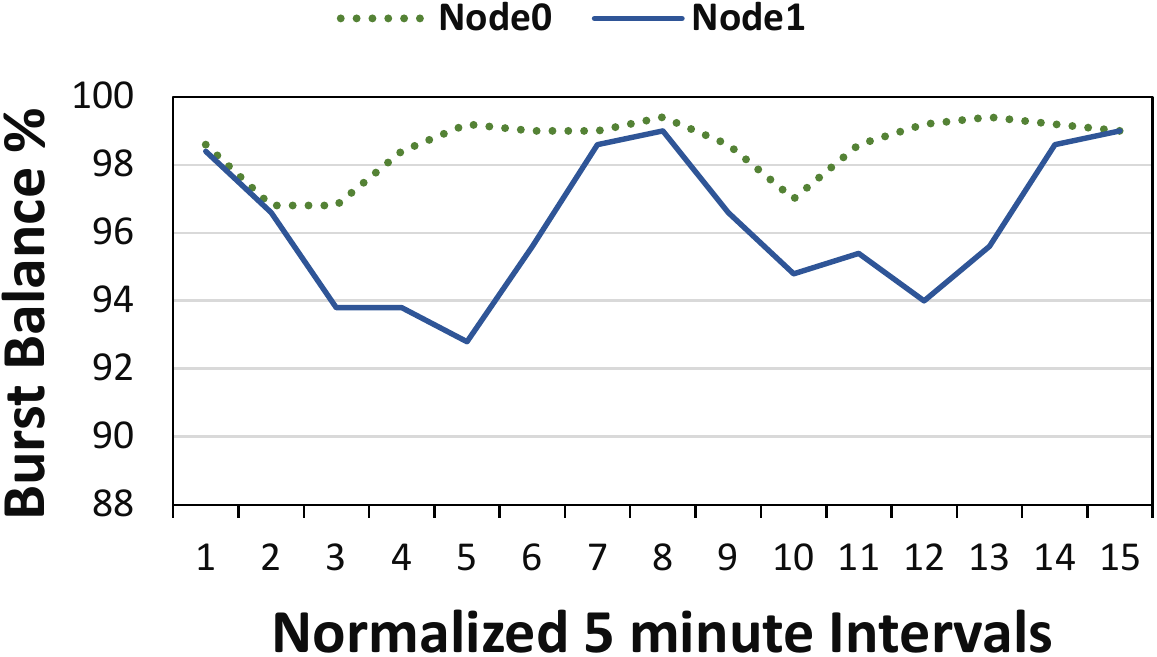}
    \caption{Disk Burst Balance on Two Nodes}
    \label{fig:diskBurstBal2Nodes}
\end{figure}

Public cloud providers maintain a credit based service rate for their network attached volume offerings as explained in section \ref{sections: background}.
Hence, it is consequential to use disk burst credits while running heavy I/O workloads.
However, cluster managers like YARN choose nodes for scheduling tasks in random order. 
A cluster manager will not differentiate between nodes which have burst credit balance and nodes which have been throttled for either CPU or disk (or network) access. 
Hence, they cannot exploit the ``burst" performance available to them in the cluster. 
This is further illustrated by a two node experiment with TPC-DS workload and observing the changes in disk burst credit in Figure \ref{fig:diskBurstBal2Nodes}. We observe a significant difference in consumption of burst credits between the two volumes attached to the two nodes in the cluster. While this does not cause any slowdown as both the disk volumes in the cluster have a full burst credit balance, we notice the uneven consumption in burst credits which has the potential for slowing down tasks if the disk volumes were running low on burst credits.

%% file: Main/Tables/t3PriceComparison.tex
\begin{table}
\begin{center}
\begin{tabular}{|l|l|l|l|}
\hline
\textbf{Instance Size} & \begin{tabular}[c]{@{}l@{}}\textbf{T3} \\ \textbf{per hour}\end{tabular} & \begin{tabular}[c]{@{}l@{}}\textbf{M5}\\ \textbf{per hour}\end{tabular} & \begin{tabular}[c]{@{}l@{}}\textbf{M5 with EMR}\\ \textbf{per hour}\end{tabular} \\ \hline
xlarge        & \$0.1664                                               & \$0.192                                               & \$0.24                                                         \\ \hline
2xlarge       & \$0.3328                                               & \$0.384                                               & \$0.48                                                         \\ \hline
\end{tabular}
\end{center}
\caption{T3 Price comparison}
\label{table:t3PriceComparison}
\end{table}

%% file: 4.methodology.tex
\section{Proposed Scheme} \label{sections: scheme}

We propose a novel scheme containing two techniques to be used together. We annotate the task requests in the 
application
frameworks to indicate the tasks for which burst credits are needed. 
We also change the cluster manager to periodically collect burst credit information from the public cloud (CPU or Disk) and use this information to match the needs of the annotated tasks.
The details of this scheduling scheme are discussed below.

\subsection{Task Annotation}

In general, CASH supports user annotations of tasks as being
intensive with respect to particular IT resources, e.g., CPU, network I/O\footnote{AWS's unorthodox dual token-bucket mechanism for network I/O of its burstable instances was reverse engineered in \cite{SIGMETRICS17}.}
and disk I/O. Users are free to experiment with associating tasks to any annotation of their choice. \par

However,
the preliminary CASH implementation
described herein relieves the user of any manual effort of annotation for the
workloads considered herein. This scheme
can either use CPU burst credits or disk burst credits but not both (i.e., one will be more of a bottleneck than the other). These ``automated" annotations happen through the framework based on the characteristics of the task's associated DAG vertex.
Specifically, DAG vertices corresponding to map-like tasks
(including ``lambda" and ``tokenize") involve the bulk of the workload processing  and so utilize resources intensively. 
Hence, these tasks can benefit the most from burst credits. Conversely, assigning these tasks to VMs which are throttled (token state is depleted) can severely affect performance. 
E.g., a workload which is I/O intensive such as database queries will need to read a large amount of data during its map-like phase and hence its map-like tasks should be assigned on VMs with high disk burst-credits. 
Similarly, workloads running on lower cost T3 burstable instances will need CPU burst credits to avoid slowdown (and heightened possibility of being
deemed stragglers).\par
The vertices of reduce-like tasks (e.g., ``reduce", ``shuffle" and ``collate") are typically less resource hungry and can be assigned on VMs where CPU/disk has been throttled. 
However, the reduce phase is generally network intensive and the framework attaches a ``network" annotation for reduce-like tasks. This leads to load balancing of network tasks in the cluster. 
The network annotation is attached along with ``CPU" or ``disk" annotation.
We observe marked improvement in reduce task execution time with MapReduce as reduce tasks are heavily network intensive. 
We provide implementation-specific details on task annotation in section \ref{sections: implementation}.

\subsection{Credit based scheduling}

\begin{figure}
    \centering
    \includegraphics[width=1\linewidth]{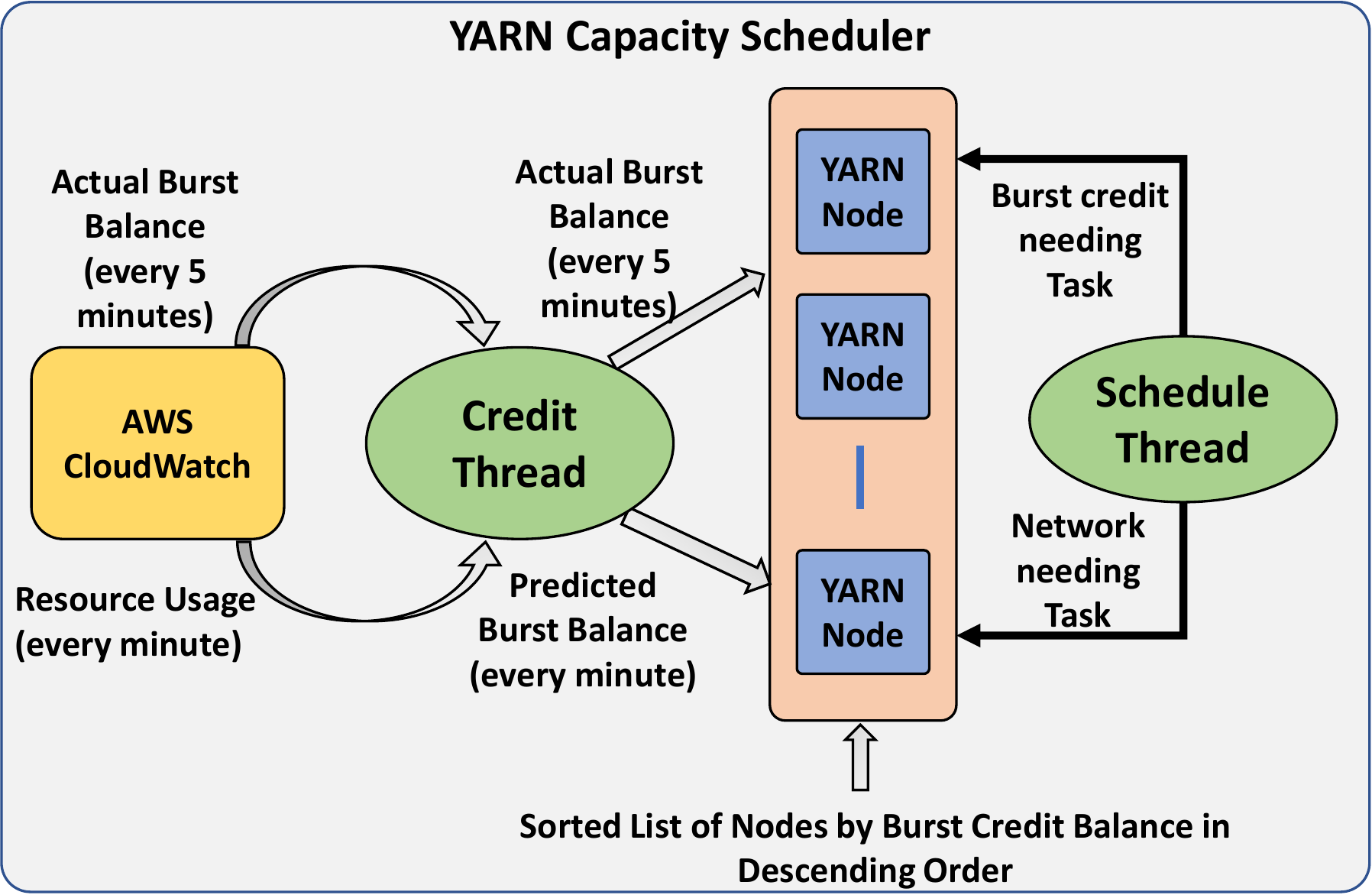}
    \caption{A schematic diagram of the proposed scheduler}
    \label{fig:design}
\end{figure}

We modify the cluster-manager's (YARN's) 
scheduler to make scheduling decisions based on burst credit balance of either the CPU or the disk volume.
Figure \ref{fig:design} shows the main components of the proposed scheduler CASH. The optimizations are described below for two cases: scheduling
 burstables (AWS T3) based on CPU credits and scheduling regular instances
 (AWS M5) based on disk I/O credits.
 In both cases, 
 different types of
 tasks of a job stream are considered.
In an ongoing work we are assessing a similar
scheduler which  jointly considers different burst credits for different
types of resources and different types of tasks (as, e.g,
rPS-DSF \cite{Shan18}), c.f. Section \ref{sections: relatedwork} for discussion of related work.
Each node has a number of slots (each corresponding to a pre-configured vCPU or virtual core) so that a node can simultaneously execute more than one task, i.e., one task per slot.
We assume the cluster manager pools all pending (annotated) tasks from all of its application frameworks into a single task queue.
Note that new tasks are continually generated by the streaming workloads handled by the application frameworks and placed in task queues and slots are freed when their tasks complete service. \par

At a long (one minute) time scale,
nodes are ordered  in decreasing order of VM CPU/disk burst credit balance
(node sorting thread).
At a short (milliseconds) time-scale, the cluster manager's scheduler (scheduler thread) first visits each node in descending order of burst credits and assigns to it as many burst (CPU or disk) intensive tasks as possible given its current number of free slots (with default optimizations like delay scheduling\cite{delay_scheduling} active), before proceeding to the next node.
This phase ends when either free node slots or burst intensive tasks in the task queue are exhausted. 
In the second phase, non-burst intensive tasks which are annotated as network are considered by the scheduling thread. 
Starting from the node with the {\em least} burst credits, in each round at most a single free slot per node is allocated to such tasks in an effort to load balance such tasks among the VMs and reduce the risk of network congestion.
Again, this phase ends either when there are no more network annotated tasks nor free slots available.
In the final phase, any remaining (non-annotated) tasks are assigned to available free slots (if any) in arbitrary node-order.
Algorithm \ref{algo: algo1} describes our credit based scheduling logic.
\input{algo1.tex} 

One pass of the scheduling thread is at the milliseconds time scale during
which new tasks may be generated by the application frameworks and slots
may become freed up. These new tasks are scheduled in the next iteration of
the scheduling thread.

%% file: algo1.tex
\begin{algorithm}
\SetAlgoLined
\DontPrintSemicolon
\While {True} {
  nodeList = sorted list of all nodes in descending order of burst credit balance\;
  \For{node in nodeList} {
     scheduleBurstIntensiveTask(node)\;
    }
    
  nodeList = sorted list of nodes in ascending order of burst credit balance\;
  \While{request in queue and slot available in nodeList} {
      \For {node in nodeList} {
        AssignOneNetworkTask(node)
      }
  }
  
  nodeList = list of nodes in random order\;
  \For{node in nodeList} {
     scheduleRemainingTask(node)\;
    }
    
   sleep(interval between scheduling)\;
 }
\caption{Schedule Thread}
\label{algo: algo1}
\end{algorithm}

%% file: 5.implementation.tex
\section{Implementation} \label{sections: implementation}
We have created a prototype of our credit based task scheduler within the YARN capacity scheduler, and conducted experiments with Tez and Hadoop applications over YARN.
We also made changes to Apache Tez to annotate tasks and communicate those annotations to YARN. We did not have to change Hadoop as we leveraged the existing node label feature of Hadoop to pass on annotations to YARN. 
All changes were made in java. We discuss our implementation in the sequel.

\subsection{YARN}
We extract the burst credit balance of each VM from Amazon Cloudwatch \cite{cloudwatch} every 5 minutes and update the internal YARN node data structure for making scheduling decisions. 
Cloudwatch populates burst credit balances at the smallest interval of 5 minutes. 
Since we do not want YARN to make scheduling decisions based on stale burst credit balance information, we also pull CPU utilization, disk read/write operations for each VM from Amazon Cloudwatch every 1 minute and predict the burst balance.
Prediction is made easy by the fact that Amazon exposes the exact formula to calculate burst credits at any given point of time based on the instance/disk size and its CPU/disk utilization. 
We update the internal YARN node data structure with predicted burst credit every 1 minute and actual burst credit every 5 minutes.
This is done in a separate asynchronous thread inside the YARN capacity scheduler. 
The process of burst credit update is described in algorithm \ref{algo: algo3}
 and is part of our design Figure \ref{fig:design}.

\input{algo3.tex}

\subsection{Tez}
\begin{figure}
    \centering
    \includegraphics[width=1\linewidth]{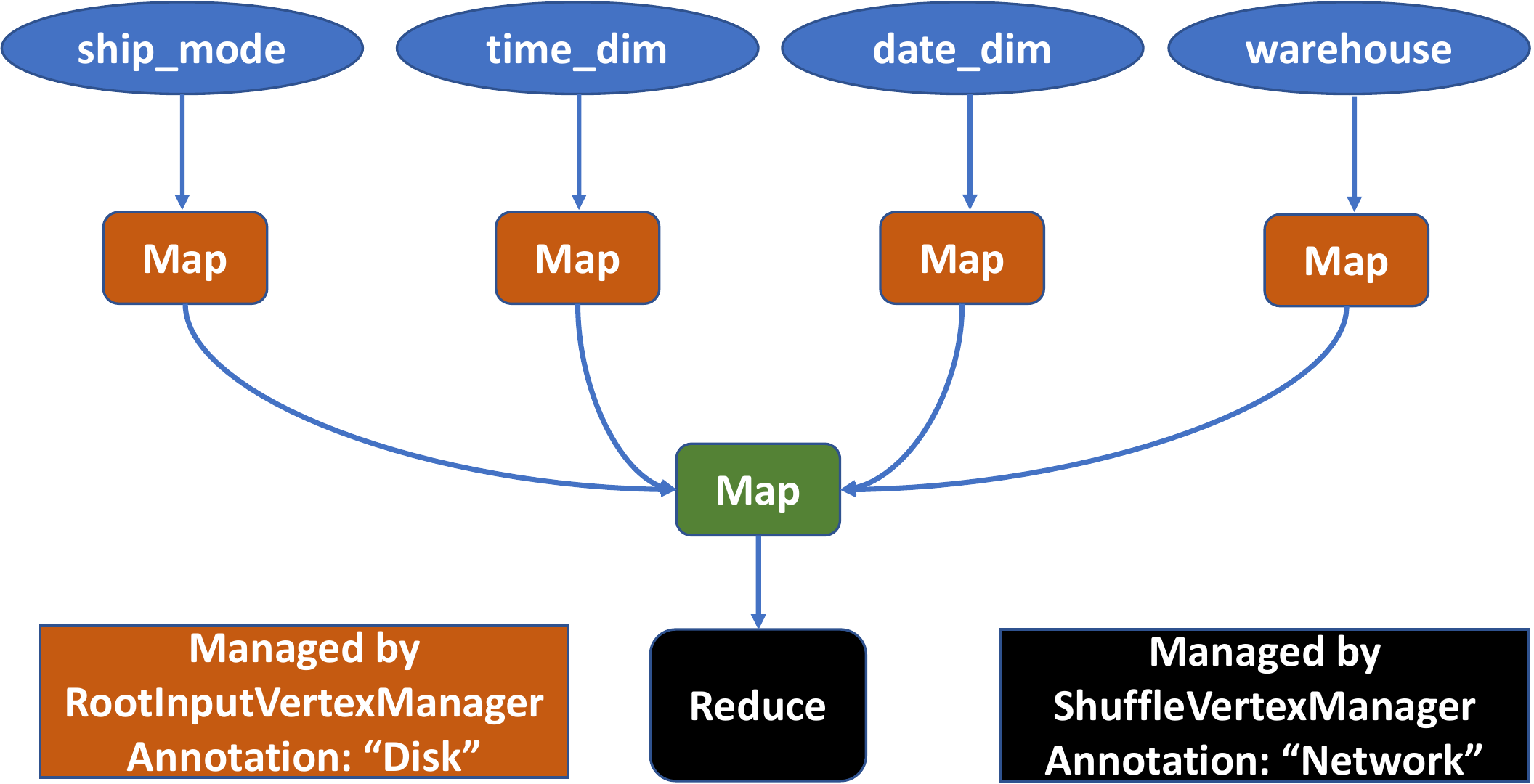}
    \caption{A Hive query DAG portion}
    \label{fig:dag}
\end{figure}

Tez provides its users with the abstraction of ``VertexManager" for dynamically adapting the execution of its tasks. Tez comes with built-in vertex managers which are associated with vertices by Tez based on the individual vertex characteristics. We modify two such vertex managers -- ``RootInputVertexManager" and ``ShuffleVertexManager" -- to implement our burst credit logic. The ``RootInputVertexManager" is associated with vertices that are the source of input data in the DAG and hence they need disk and/or CPU  burst credits. Similarly, ``ShuffleVertexManager" is associated with vertices that shuffle network data and can benefit from better network load balancing. These vertex managers annotate the task requests to YARN for their associated  vertices. 
We provide a portion of an actual Hive query execution DAG with their associated vertex managers and annotations in Figure \ref{fig:dag}.\par

Tez allows its users to create their own vertex managers and associate them explicitly with the vertices of the DAG they create (unlike Spark, Tez programmers can create their own DAG). We have added the annotation feature to the base level class of vertex manager in Tez source code allowing users to associate annotation with their user defined vertex managers. This allows users to associate any annotation with any vertex of their choice in their execution DAG.\par

\subsection{Hadoop}
Every Hadoop job can be expressed as a DAG with two vertices - ``Map" and ``Reduce". We associate burst annotation (CPU or disk based on user configuration) with map vertex and network annotation to reduce vertex. This allows resource hungry map tasks to be assigned burst credits and (generally) network heavy reduce tasks to be well-balanced.

%% file: algo3.tex
\begin{algorithm}
\SetAlgoLined
\DontPrintSemicolon
\While {True} {
    \If {timeInterval() == 5 minutes} {
        \If {Annotation == CPU credit} {
            getCPUBurstCreditsFromCloudWatch()\;
        }
        \Else{
            getDiskBurstCreditsFromCloudWatch()\;
        }

        setBurstCreditsOnAllNodes()\;
    } 
    \If {timeInterval() == 1 minute} {
        \If {Annotation == CPU credit} {
            getCPUUsageFromCloudWatch()\;
        }
        \Else{
            getDiskIOUsageFromCloudWatch()
        }
            
        setCalculatedBurstCreditsOnAllNodes()\;
    }
}
\caption{Burst Credit Fetch Thread}
\label{algo: algo3}
\end{algorithm}

%% file: 6.evaluation.tex
\section{Evaluation} \label{sections: evaluation}
\begin{figure}
    \centering
    \includegraphics[width=1\linewidth]{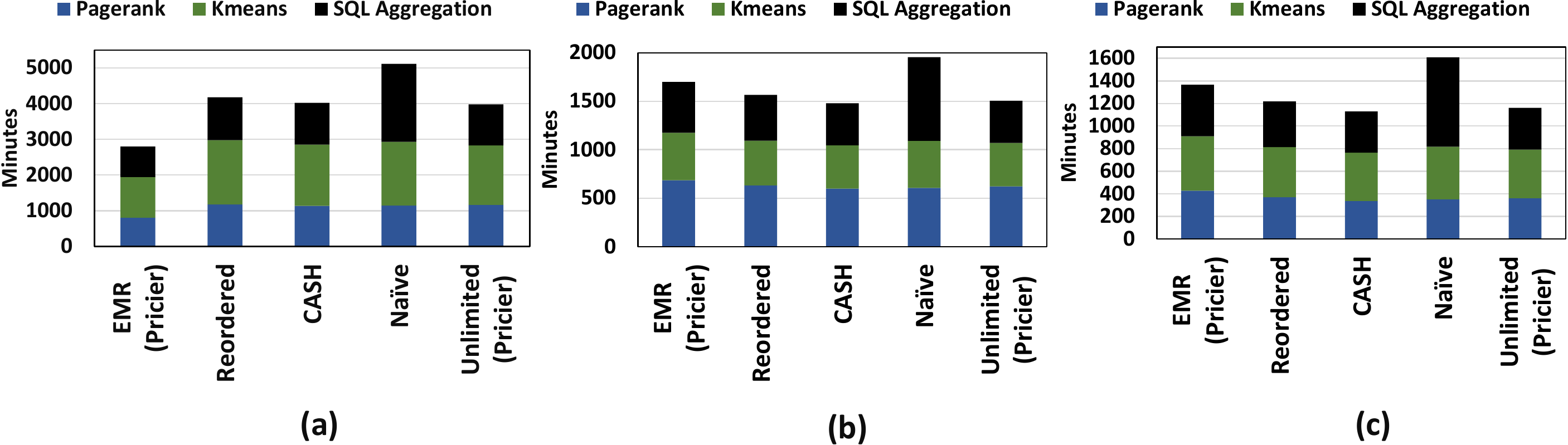}
    \caption{Cumulative elapsed time: (a) Map, (b) Reduce, (c) Shuffle}
    \label{fig:cumElapsedTime}
\end{figure}
\begin{figure}
     \centering
        \includegraphics[width=1\linewidth]{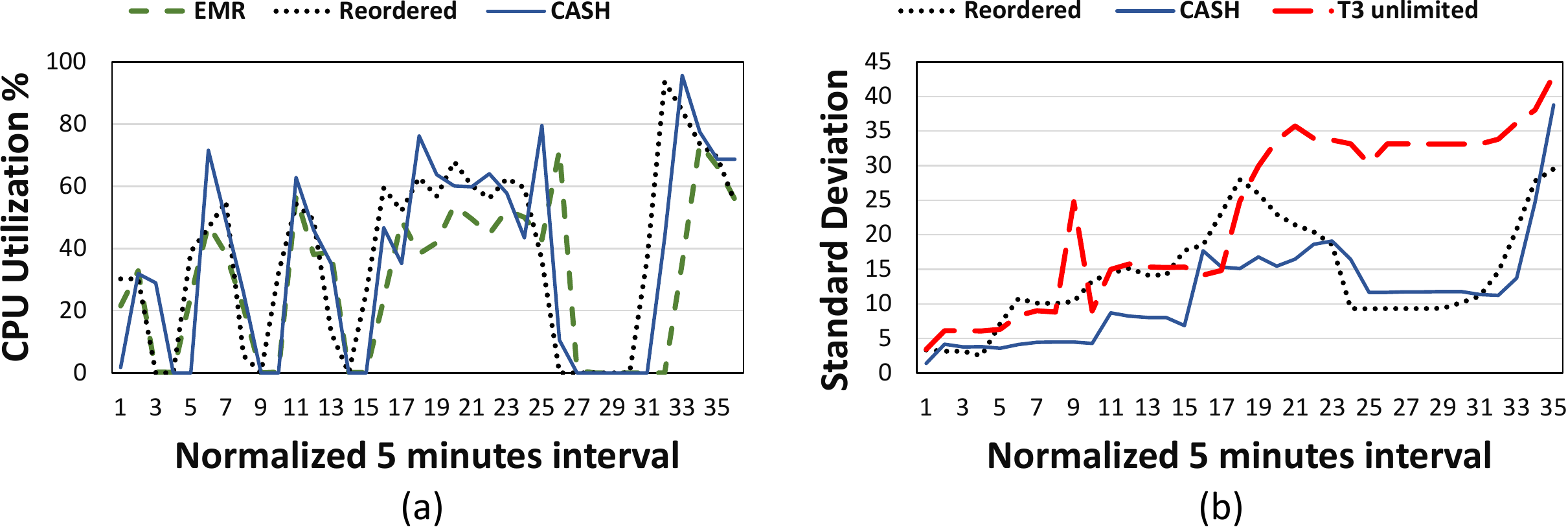}
    \caption{(a) Average CPU Utilization Timeline, (b) Standard deviation of CPU Credit Balance }
    \label{fig:cpuUtilstdDevCpuCredit}
\end{figure}
\subsection{Workloads -- CPU Burst}
We use Intel HiBench \cite{hibench} to run batch workloads for CPU burst evaluation. It provides several workloads, including machine learning, graph processing and Hive SQL queries, among others. We use the workloads PageRank, K-means clustering and Hive SQL aggregation for our evaluation. We choose these workloads as they are representative of popular Hadoop batch workloads. We generate the input data for these workloads synthetically through HiBench itself. The generated data is written to S3 and is used as the input data for all experimental runs. Each HiBench workload comprises of several jobs. These jobs are submitted sequentially, with the input of a job being dependent on the output of the job prior to it. All jobs read their input data from S3 and write their output data to S3. A job submitted to YARN is broken down into smaller tasks which have their own dependency graph and use the local VM storage(EBS) as temporary scratch space.

\subsection{CPU Burst: Experiment Design and Setup}
We generate our experimental baseline using Amazon EMR. EMR comes pre-configured with Hadoop and is highly optimized to work with S3. We do not change any configuration in EMR and run the workloads sequentially.
Using EMR to create our baseline removes any biases that we might inadvertently introduce into it. We use EMR version 5.28 which is based on open source Hadoop 2.8.5. 

To run CASH, we install Hadoop 2.8.5 over 10 EC2 T3.2xlarge instances which has a baseline CPU throughput of 40\%.
We copy most of the compatible Hadoop configuration (such as slot size, map memory, reduce memory, node manager memory, compression algorithm, etc.) from EMR to our T3 installation so as to keep the configurations similar.
Again, our proposed scheduler is prototyped on the YARN capacity scheduler, with most configurations retained from EMR. Details of experiment are in sequel.

\subsubsection{CPU Burst Experiment-1: Naive workload submission}
We run SQL aggregation, which has an average CPU requirement greater than the baseline CPU throughput of the AWS instance as our first workload. The CPU is throttled to 40\% as the instances have not accumulated any burst credits to support higher than 40\% CPU. We run PageRank and K-means after SQL aggregation. 

\subsubsection{CPU Burst Experiment-2: Reordered workload submission}
We run workloads in the order PageRank, K-means and SQL aggregation. This allows credits to be accumulated in the beginning through the first two workloads that operate at lower than baseline CPU throughput. Accumulating credits in the beginning allows SQL aggregation to be executed without being throttled due to lack of CPU credits. This effort can be easily automated by pulling job CPU utilization metrics from a historical metric server of choice, such as Amazon Cloudwatch \cite{cloudwatch} or the Hadoop history server \cite{historyServer}. Tenants who do not have any workloads that have CPU needs above the baseline CPU utilization may submit their workloads in any order of their choice.

\subsubsection{CPU Burst Experiment-3: T3 unlimited}
In this experiment, we run workloads with T3 unlimited mode ON. With T3 unlimited, an instance is never throttled when it uses up all its burst credits. The CPU credits are averaged over a period of 24 hours and a tenant is billed for the burst credits they use above the baseline rate. A T3 unlimited instance using 52.5\% average CPU over 24 hours will be billed the same as an equivalent general purpose instance (regular VM), and a T3 instance using 100\% CPU over 24 hours will be billed close to 50\% more than a comparable general purpose instance. This option is exclusive to AWS and tenants of providers like Azure do not yet have this option.

\subsubsection{CPU Burst Experiment-4: Workload submission with CASH}
We change YARN capacity scheduler to account for CPU burst credits and network intensive tasks while scheduling. With the 
CPU burst credit and network aware YARN (a.k.a. CASH), workloads 
are submitted so that CPU-intensive workloads are submitted last.
\begin{figure}
    \centering
    \includegraphics[width=1\linewidth]{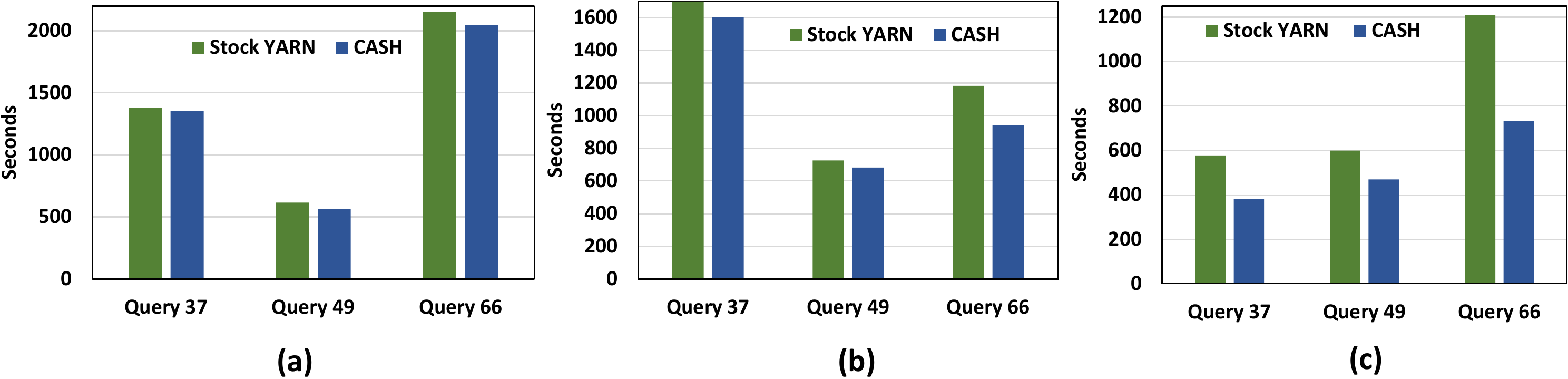}
    \caption{Query Completion Time Comparison (a) 2 VMs, (b) 10 VMs, (c) 20 VMs}
    \label{fig:queryTimeComparison}
\end{figure}

\subsection{Results -- CPU Burst}
We report our results through a cumulative elapsed time comparison of the three phases a job goes through in Hadoop namely -- map, shuffle and reduce. 
We do not report the overall makespan of our workloads as access to AWS S3 storage varies widely over time \cite{pelleS3Latency}. This may be due to 
dynamic demand or due to an undocumented token-bucket mechanism or both. For this reason,
in the experimental results that follow, we report component task execution-times
rather than overall workload wall-clock (makespan) execution times. Also, considering EMR is SaaS, we expect that
generally, S3 variation will be larger for non-SaaS implementations 
proposed herein.\par
Figure \ref{fig:cumElapsedTime} provides the results observed.
Running workloads over T3 naively resulted in a 40\% degradation in elapsed time resulting in T3 being more expensive than EMR.
CASH performs better than naive and reordered workload submission in all phases and degrades by about 13\% in cumulative task elapsed time compared to EMR. However, running T3 is about 30.7\% cheaper than running EMR and hence tenants will save cost by running their workloads on T3 using CASH. A 13\% degradation in cumulative elapsed task time doesn't translate to equivalent degradation in makespan time as the phases map, reduce and shuffle overlap with each other. Particularly, the reduce stage starts as early as when 5\% of the mapper output is available by starting to shuffle the map task outputs. Running workloads over T3 by simple reordering of workloads leads to a degradation of about 19\% compared to EMR which doesn't translate to as much cash savings as CASH. T3 instances running CASH also show better CPU utilization than simple reordering and EMR pointing to better load balancing as shown in Figure \ref{fig:cpuUtilstdDevCpuCredit}(a).\par

Running workloads on T3 with unlimited option ON yields about the same elapsed time as CASH but the former has caveats. T3 unlimited averages CPU utilization on a per instance basis. Tasks can be scheduled on VMs which have zero credit balance causing them to be billed for additional credits. This is possible while there are other VMs present in the cluster with surplus credits. Hence, tenants will be billed for additional credits while there are surplus credits available in the cluster. This phenomenon can be ascertained by observing the standard deviation of CPU credit balance across all VMs of the cluster in Figure \ref{fig:cpuUtilstdDevCpuCredit}(b). With the high standard deviation of T3 unlimited, tenants are billed extra for excess CPU credits which could have been avoided. Hence, CASH delivers more cash compared to T3 unlimited.

\subsection{Workloads -- Disk Burst}
We use hive-testbench \cite{hive-testbench} to run streaming queries for disk burst evaluation. Hive-testbench is a benchmark suite based on industry standard TPC-DS queries to test database systems. We use three TPC-DS queries to be run over Hive database using Tez -- query 66, query 49 and query 37. We choose the queries which read a high amount of data from disk to create scenarios where disk credits are significantly depleted.
Input data is generated by hive-testbench and is stored in HDFS as a hive warehouse. We use Hive-2.3.6 with Tez-0.9.2 over YARN-2.8.5, the equivalent versions used in EMR.

\subsection{Disk Burst: Experiment Design and setup}\label{subsection: disk_burst_exp}
We run all three TPC-DS queries in parallel. To avoid bias due to data caching, we have disabled query caching in Hive. 
We restart the instances of the cluster between consecutive experiments and write random data on the disk volumes so as to invalidate the disk cache. We test our optimization gains by comparing execution time and wall-clock completion time against running the same queries on stock YARN keeping the same cluster and the database. \par

At the beginning of each experiment, we wipe out the disk credits and start with zero burst credits. We do this for two reasons: (i) Amazon SSD volumes come with 5.4 million startup burst credits which is not a realistic burst credit balance to expect at a typical point in time in a long running cluster.  (ii) We want to explore the scenario in which instance volumes run out of burst credits, potentially leading to task slowdown, and how this threat can be addressed.

\subsubsection{Disk Burst: Experiment-1 Two VMs}
We run our experiment on two EC2 M5.2xlarge instances with hive database size of 280 GB and EBS volume size of 200 GB per instance.

\subsubsection{Disk Burst: Experiment-2 Ten VMs}
We run our experiment on ten EC2 M5.2xlarge instances with hive database size of 1.2 TB and EBS volume size of 170 GB per instance.

\subsubsection{Disk Burst: Experiment-3 Twenty VMs}
We run our experiment on twenty EC2 M5.2xlarge instances with hive database size of 2.5 TB and EBS volume size of 200 GB per instance.

\subsection{Results -- Disk Burst}
We calculate the wall-clock time (makespan) as the time it takes for all three queries to have returned their output, while we consider query completion time as the time taken for a particular query to return its output.
We observe an average improvement of about 5\% in query completion time and overall wall-clock time improvement of 4.85\% in our two VM experiment with 280 GB input database. The results are in Figure \ref{fig:queryTimeComparison}(a).
The improvements are modest as the query IOPS requirement on a 280 GB database is low. The tasks spawned for query execution needed few burst credits.\par

\begin{figure}
    \centering
    \includegraphics[width=1\linewidth]{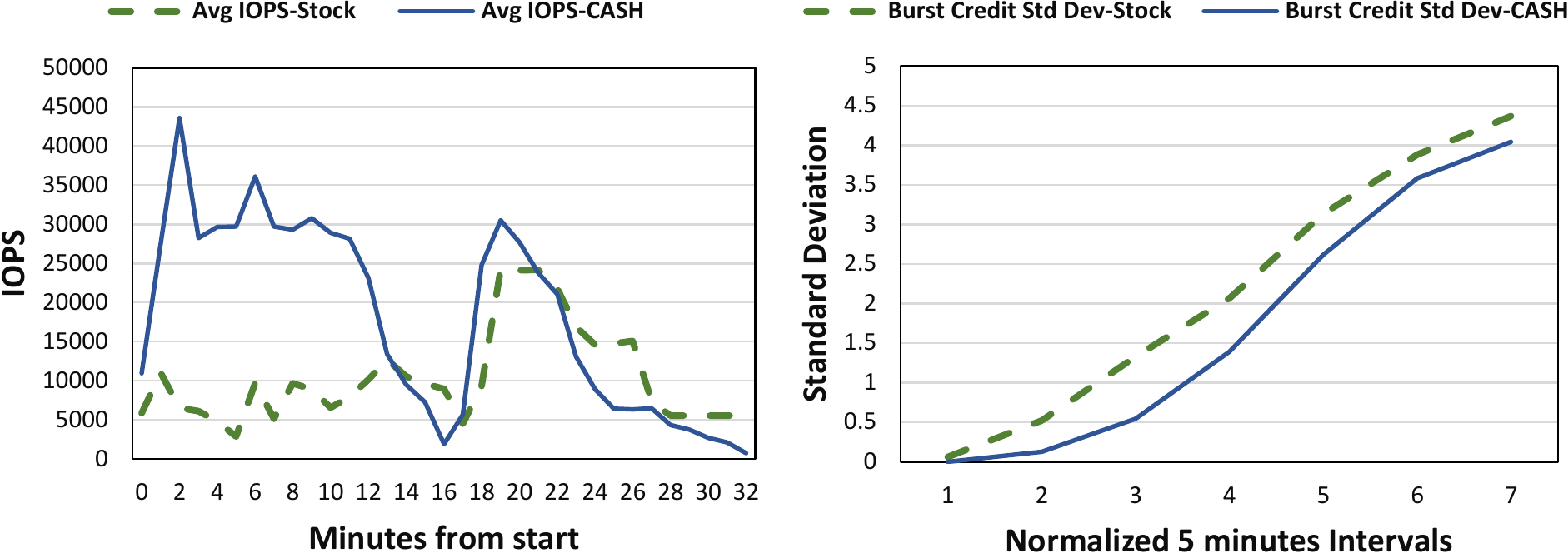}
    \caption{(a) Avg Total IOPS, (b) Std Deviation of Burst Credits in the cluster}
    \label{fig:IOPSstdDevcomparison10nodes}
\end{figure}

In the experiment with 10 VMs and an input database of 1.2 TB, the IOPS requirement of the queries were much higher even with a 5 times increase in compute capacity.
The cluster running CASH showed much higher average IOPS and a lower standard deviation of disk burst credit balance across the volumes of the cluster compared to stock YARN as shown in Figure \ref{fig:IOPSstdDevcomparison10nodes}. High average IOPS leads to less I/O wait time in task execution. A lower standard deviation of burst credit balance points to better load balancing of I/O tasks in the cluster. CASH was able to opportunistically schedule I/O bound tasks on VMs with higher disk burst credit balance leading to peaks in I/O.
CASH running on a 1.2 TB database improved average query completion time by about 10.7\% and overall wall-clock time by about 13\% compared to stock YARN. Results in Figure \ref{fig:queryTimeComparison}(b).\par
We hypothesize, the more I/O-intensive a workload is, the more speedup CASH can provide. In order to test this hypothesis, we ran several  experiments with larger database sizes and we report one such experiment as experiment three. With an increased scale of 2.5 TB database, CASH improved average query completion time by 31\% and wall-clock time by 22\%. Results are in Figure \ref{fig:queryTimeComparison}(c).\par

\begin{figure}
    \centering
    \includegraphics[width=1\linewidth]{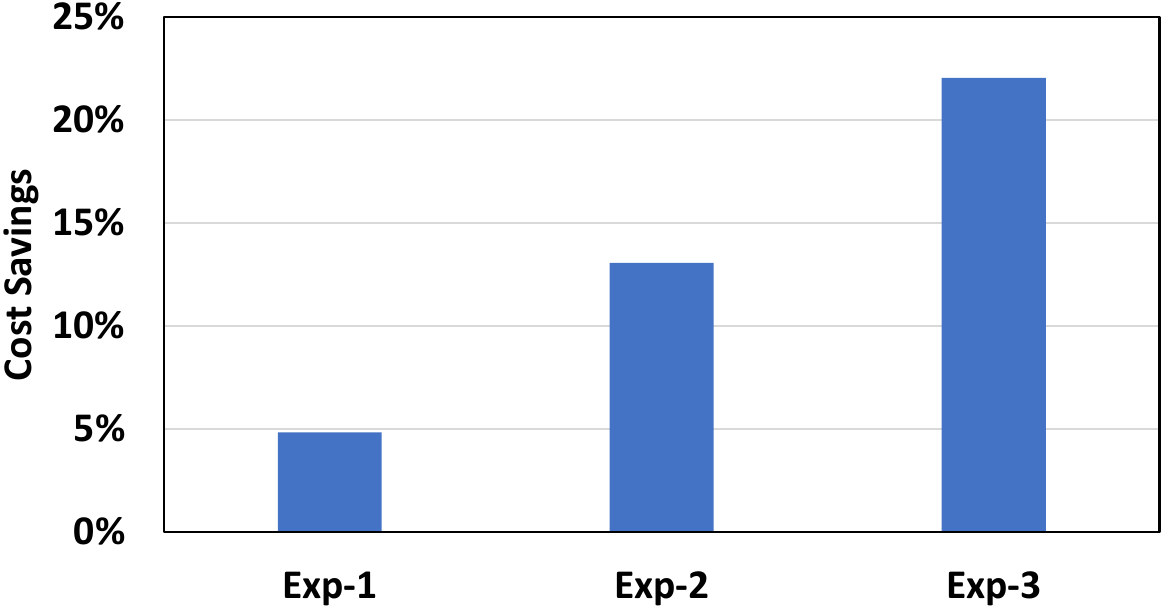}
    \caption{Cost savings through CASH (Experiment details in section \ref{subsection: disk_burst_exp})}
    \label{fig:CASHcostSavings}
\end{figure}

Any improvement in end-to-end wall-clock time directly translates to cost savings of equal valuation in terms of public cloud billing.
Hence, our wall clock time improvement of upto 22\% directly translates to a costs savings of equal amount for tenants of the public cloud using CASH. We summarize our cost savings in Figure \ref{fig:CASHcostSavings}.

%% file: 7.relatedwork.tex
\section{Related Work} \label{sections: relatedwork}

Prior work such as \cite{wangSpotBurstable} show that burstable instances can be used as a passive backup system which is also highly available.
\cite{SIGMETRICS17} characterizes the unorthodox dual token-bucket mechanism used by AWS burstables and proposes some use cases. 
However, this work was done in conjunction with spot instances over memcached workloads. 
\cite{burstingPossibilities} gives an empirical study on burstable instances and two use cases. 
They suggest that burstable instances be used for applications with low or irregular CPU loads, hosting non critical I/O or network bound jobs or to simply restart the instances to reclaim the launch credits. 
\cite{burstableModelling} create a theoretical model to maximize revenue from burstable instances from a provider perspective. 
Tenants can also use their model to choose the instance size of their burstable instances for minimizing costs.
More recently, \cite{HeMT-arxiv} dynamically {\em resizes} task slots for burstable instances based on their current CPU token state and finally \cite{burscale} shows how to autoscale using burstables.

None of these works except \cite{HeMT-arxiv}, look at a variable service rate VM like the burstables for running large scale data processing frameworks and even \cite{HeMT-arxiv} looks at only CPU credits for their optimizations. To the best of our knowledge, this is the first such work in this field which considers the variable service rate of hardware resources and looks at burst credit balance of CPU and disk (separately) for making optimal placement decisions.

This said, the scheduling problems considered in this paper are not unrelated to rPS-DSF \cite{Shan18}, based on  PS-DSF \cite{Jalal18}, which generalizes scheduling based on Dominant Resource Fairness (DRF) criteria \cite{DRF}. 
Papers related to DRF typically assume statically available resources (as would not be the case for burstable instances) and that more detailed resource needs of each task are known. 
A completely different approach to large scale data processing \cite{monotasks} is well suited for running workloads on a credit based public cloud scheduler such as CASH. This approach breaks a job into tasks s.t. each task uses a single hardware resource.

%% file: 8.conclusion.tex
\section{Summary and Future Work} \label{sections: summary}

We propose a novel scheduler for cluster VMs, particularly burstable VMs, of a public
cloud which relies on the knowledge of the burst
credit/token states of the IT resources
of each VM and, only roughly,
 the  resource needs of the tasks. 
 The basic idea is to avoid assigning tasks which employ a particular IT resource intensively (e.g., CPU or disk I/O) to VMs which are currently low on tokens for that resource. Token state is obtained from the public cloud, while such resource needs of tasks are communicated by the application (or application framework). The scheduler was prototyped on YARN and experiments were conducted on Tez and Hadoop over YARN with different workloads. We were able to run low CPU-intensive Hadoop workloads over T3 burstable instances while lowering costs. We were also able to accelerate streaming hive queries through our optimization and save costs by 22\%.
In on-going work, we are experimenting with {\em joint} scheduling of plural credit-based resources (CPU, disk I/O and network I/O).